\documentclass[prl,twocolumn,showpacs,preprintnumbers,amsmath,amssymb,floats,citeautoscript,nobalancelastpage]{revtex4-1}
\usepackage{graphicx}
\usepackage{dcolumn}
\usepackage{bm}
\usepackage{txfonts}
\usepackage{color}

\begin{document}

\title{Optical Conductivity Measurement of a Dimer-Mott to Charge-Order Phase 
Transition in a Two-Dimensional Quarter-Filled Organic Salt Compound}

\author{Ryuji~Okazaki$^{1,\ast}$}
\author{Yuka~Ikemoto$^2$}
\author{Taro~Moriwaki$^2$}
\author{Takahisa~Shikama$^{3}$}
\author{Kazuyuki~Takahashi$^{3,4}$}
\author{Hatsumi~Mori$^{3}$}
\author{Hideki~Nakaya$^{5}$}
\author{Takahiko~Sasaki$^{5}$}
\author{Yukio~Yasui$^{1,6}$}
\author{Ichiro~Terasaki$^{1}$}

\affiliation{$^1$Department of Physics, Nagoya University, Nagoya 464-8602, Japan}
\affiliation{$^2$SPring-8, Japan Synchrotron Radiation Research Institute (JASRI), Sayo, Hyogo 679-5198, Japan}
\affiliation{$^3$Institute for Solid State Physics, The University of Tokyo, Kashiwa, Chiba 277-8581, Japan}
\affiliation{$^4$Department of Chemistry, Kobe University, Kobe 657-8501, Japan}
\affiliation{$^5$Institute for Materials Research, Tohoku University, Sendai 980-8577, Japan}
\affiliation{$^6$Department of Physics, Meiji University, Kawasaki 214-8571, Japan}

\date{\today}

\begin{abstract}
We report a novel insulator-insulator transition 
arising from the internal charge degrees of freedom 
in the two-dimensional quarter-filled organic salt $\beta$-({\it meso}-DMBEDT-TTF)$_2$PF$_6$.
The optical conductivity spectra above $T_c = 70$~K
display a prominent feature of the dimer-Mott insulator,
characterized by a substantial growth of a dimer peak near 0.6~eV with decreasing temperature.
The dimer-peak growth is rapidly quenched as soon as 
a peak of the charge order shows up below $T_c$,
indicating a competition between the two insulating phases.
Our infrared imaging spectroscopy has further revealed 
a spatially competitive electronic phases far below $T_c$,
suggesting a nature of quantum phase transition 
driven by  material-parameter variations.
\end{abstract}



\maketitle

Organic molecular conductors exhibit complex electronic phase diagram 
owing to their unique internal degrees of freedom coupled with correlation effects.
Among them, the quasi two-dimensional (2D) quarter-filled salts $R_2X$, 
where $R_2$ is a dimer organic molecule and $X$ is a monovalent anion,
are attracting much interest \cite{Seo06,Hotta12}.
In the weakly dimerized materials,
the correlated carrier is localized on the molecular site due to 
long-range nature of Coulomb interaction, 
leading to charge order as seen in $\theta$-(ET)$_2$RbZn(SCN)$_4$ \cite{Miyagawa00} and 
$\alpha$-(ET)$_2$I$_3$ \cite{Takano01}
[ET=bis(ethylenedithio)-tetrathiafulvalene].
On the other hand, under strong dimerization,
the hole is localized on the dimer 
to act as a Mott insulator (the dimer-Mott insulator),
as realized in $\kappa$-(ET)$_2X$, where $X$=Cu$_2$(CN)$_3$, Cu[N(CN)$_2$]Cl.

Recently, intriguing behaviors have been found in
materials located near the border of such insulating phases.
In the dimer-Mott insulator $\kappa$-(ET)$_2$Cu$_2$(CN)$_3$, 
the localized hole behaves as a magnetic dipole but shows no
magnetic order down to very low temperatures,
leading to a novel spin-liquid state \cite{Shimizu03,SYamashita08,MYamashita09}.
Interestingly,
in sharply contrast to conventional Mott insulators,
this dimer Mott insulator exhibits unusual temperature and frequency 
variations of dielectric constant \cite{Abdel}.
This experimental fact strongly suggests that the ET dimer is \textit{electrically} polarized 
owing to the intra-dimer charge degree of freedom,
which Hotta called ``dipolar liquid'' \cite{Hotta10}.
The dielectric anomaly is also observed in the dimer-Mott insulators
$\kappa$-(ET)$_2$Cu[N(CN)$_2$]Cl \cite{kCl} and
$\beta '$-(ET)$_2$ICl$_2$ \cite{bICl}.
These results
have revealed that the instability to electric dipole order (charge order)
is hidden in the dimer-Mott insulators \cite{Naka10}.
The collective mode of the electric dipole order 
is theoretically calculated to be gapless
or extremely narrow-gapped \cite{gapless,Naka12}, and can affect the 
low-temperature specific heat and thermal conductivity \cite{SYamashita08,MYamashita09}. 
On the other hand, in these dimer-Mott insulators, 
there is no direct spectroscopic evidence of charge disproportionation \cite{Shimizu06, Sed12,Tom12},
which can be sensitively probed through 
optical conductivity \cite{rev04,rev12} or nuclear magnetic resonance measurements \cite{Takahashi06}.
These results indicate that the observed dielectric anomaly is not simply attributed to charge order,
and then raise a crucial question whether 
the instability toward the charge order exists in such dimer-Mott phase.
The aim of this paper is to explore the opposite case  
where the instability to the dimer-Mott phase is
hidden in a well established charge-ordered dimer-type organic conductor.  
Here we show such a phenomenon in the dimer-type quarter-filled
organic salt $\beta$-({\it meso}-DMBEDT-TTF)$_2$PF$_6$ 
through the infrared optical study including a spatial imaging measurement.

The polarized reflectivity spectra in a large area ($\sim 400 \times50$ $\mu$m$^2$) on the sample surface 
were measured for energies between 90 meV and 1.4 eV using a 
Fourier transform infrared spectrometer (FTIR) equipped with an infrared microscope.
The details of measurement area will be shown later.
The sample was fixed with a conductive carbon paste on the cold head of helium flow-type refrigerator. 
The sample cooling rate was about 1~K/min.
We used a standard gold overcoating technique  for measuring the reference spectrum at each temperature.
The complex optical conductivity is obtained from the Kramers-Kronig (KK) analysis.
Standard extrapolation of $\omega^{-4}$ dependence was employed above 1.4 eV.
In low energies, we extrapolated the reflectivity using several methods,
but the peak shapes and positions are negligibly affected.
Infrared-imaging measurements using a synchrotron radiation (SR) light were performed at BL43IR, SPring-8, Japan \cite{HKimura04,SKimura04}.
We measured the positional dependence of the $c^*$-axis-polarized local reflectivity spectra  on a crystal surface of $1400\times 100$ $\mu$m$^2$ by using an FTIR for energies between 0.1 eV and 0.8 eV.
High spatial resolution of $\sim$ 10 $\mu$m 
was achieved with high-brilliance SR light (spot diameter of 10 $\mu$m) and 
an infrared microscope equipped with a precision {\it xy}-scanning stage.

\begin{figure}[t]
\includegraphics[width=1\linewidth]{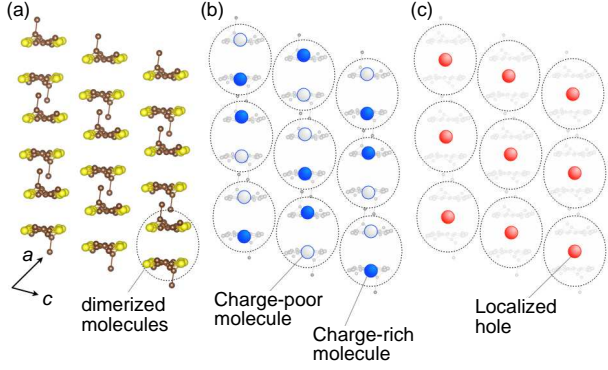}
\caption{(Color online) 
(a) The two-dimensional (2D) molecular plane of $\beta$-({\it meso}-DMBEDT-TTF)$_2$PF$_6$, 
where DMBEDT-TTF = 2-(5,6-dihydro-1,3-dithiolo[4,5-{\it b}][1,4]dithiin-2-ylidene)-5,6-dihydro-5,6-dimethyl-1,3-dithiolo[4,5-{\it b}][1,4]dithiin.
Hydrogen atoms are omitted for clarity. Two donor molecules circled by the dotted ellipsoid are dimerized. 
(b) A checkerboard-type charge-ordered state below $T_c = 70$ K.
(c) A dimer-Mott insulating state above $T_c$. 
}
\end{figure}

The single crystals of $\beta$-({\it meso}-DMBEDT-TTF)$_2$PF$_6$ were grown by 
the electrochemical method \cite{KimuraCC},
where DMBEDT-TTF stands for 2-(5,6-dihydro-1,3-dithiolo[4,5-{\it b}][1,4]dithiin-2-ylidene)-5,6-dihydro-5,6-dimethyl-1,3-dithiolo[4,5-{\it b}][1,4]dithiin.
The crystal structure is composed of the
stacking of conducting donor layers
separated by insulating anionic ones \cite{KimuraCC}.
In the conducting layer, two donor molecules are weakly dimerized 
as shown in Fig. 1(a).
The x-ray diffraction \cite{KimuraJACS} as well 
as the infrared and Raman studies \cite{Tanaka08} clearly resolve 
a charge disproportionation below $T_c = 70$~K.
The charge ordering pattern 
is of checkerboard type [Fig. 1(b)] \cite{KimuraJACS},
which can be regarded as an antiferro-type electric dipole order,
contrast to stripe-type charge ordered states
in other 2D quarter-filled  salts
\cite{Takahashi06}.

\begin{figure}[t]
\includegraphics[width=1\linewidth]{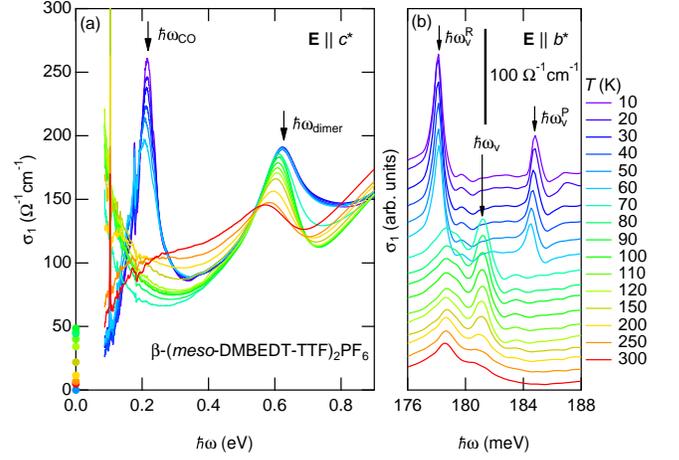}
\caption{(Color) 
The real part of optical conductivity $\sigma_1$ of $\beta$-({\it meso}-DMBEDT-TTF)$_2$PF$_6$ single crystal measured in the large area on the sample surface. 
(a) Temperature-dependent $\sigma_1$ spectra measured with polarization parallel to the {\it c}* axis. 
The dc conductivity $\sigma_1(\omega \to 0)$ is plotted with the circles.
(b) Expanded view of a temperature-dependent vibrational molecular modes around 180 meV in $\sigma_1$ spectra measured with polarization parallel to the {\it b}* axis.}
\end{figure}

Figure 2(a) displays the real part of the optical conductivity 
spectra $\sigma_1$  obtained from the KK transformation 
of the reflectivity spectra measured in the large area on the sample surface 
with polarization parallel to the conducting {\it c}* axis.
A low-energy spectrum below 0.2 eV 
is gradually enhanced with decreasing temperature above $T_c$
but is suddenly suppressed and replaced 
by a sharp peak structure at $\hbar\omega_{\rm CO} \simeq$ 0.2 eV at $T_c$.
This indicates that the formation of charge order
drastically modifies the electronic structure  in a wide energy range and 
such spectroscopic feature has also been observed in other charge-ordered material \cite{Ivek11}.
The vibrational molecular modes in $\sigma_1$ measured 
with polarization parallel to the insulating {\it b}* axis 
provide evidence for charge order below $T_c$ [Fig. 2(b)],
consistent with previous results \cite{Tanaka08}.
As shown in Fig. 2(b), the $\nu_{14}$ mode peak
(out-of-phase stretching mode of ring C=C) of DMBEDT-TTF$^{0.5+}$ 
at $\hbar\omega_{\rm v} \simeq181$ meV splits into two bands 
(charge-rich site at $\hbar\omega_{\rm v}^{\rm R} \simeq178$ meV 
and charge-poor site $\hbar\omega_{\rm v}^{\rm P} \simeq185$ meV) 
owing to the charge disproportionation below $T_c$.

Firstly we discuss the high-temperature phase above $T_c$.
The distinct feature newly found in $\sigma_1$
is a pronounced peak structure at $\hbar\omega_{\rm dimer} \sim 0.6$~eV,
which exhibits a strong enhancement 
with lowering temperature from 300 K down to $T_c$.
This mid-infrared peak can be assigned to a dimer peak, 
a transition from bonding to anti-bonding orbitals 
of dimerized molecules, as observed in $\kappa$-type dimerized ET salts \cite{Faltermeier07}.
Most importantly, the dimer-peak intensity is enhanced 
with decreasing temperature in the dimer-Mott insulating phase, 
while it is reduced in a correlated metallic phase \cite{Sasaki04}.
Thus the observed enhancement of  dimer-peak intensity down to $T_c$ strongly
indicates that the high-temperature phase in this material should be
regarded as a dimer-Mott insulating phase [Fig. 1(c)], 
rather than a conventional metal.
Below 200 K, $\sigma_1$ seems to exhibit a Drude-like response below 0.3 eV, 
but it should show a peak structure centered at a finite energy to connect low $\sigma_{\rm DC}$ values plotted in Fig.~2(a).
Such a low-energy response strikingly resembles those of the quarter-filled salt $\theta$-(ET)$_2$I$_3$, 
in which an incoherent transport is realized at high temperatures \cite{Takenaka05}.

\begin{figure}[t]
\includegraphics[width=1\linewidth]{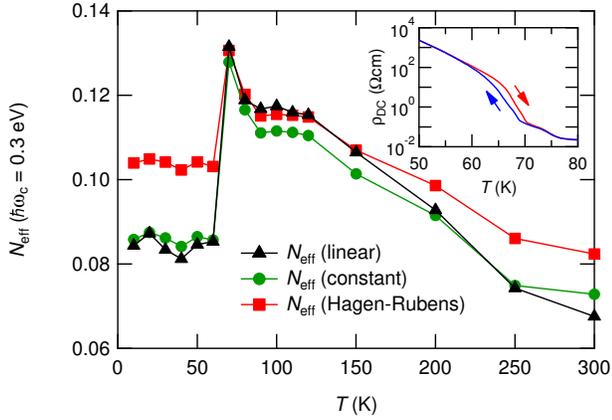}
\caption{
(Color online)
Temperature dependence of the effective carrier number $N_{\rm eff}$ per formula unit calculated up to 0.3 eV.
Several results obtained using different low-energy extrapolations are displayed.
Inset: temperature variation of dc resistivity $\rho_{\rm DC}$ near $T_c$.
}
\end{figure}

To shed further light on the unusual electronic state in
high-temperature dimer-Mott phase of $\beta$-({\it meso}-DMBEDT-TTF)$_2$PF$_6$,
we discuss the effective carrier number $N_{\rm eff}$ expressed by, 
\begin{equation}
N_{\rm eff}(\omega_c) = \frac{2m_0V}{\pi e^2}\int_0^{\omega_c}\sigma_1(\omega')d\omega',
\end{equation}
where $m_0$ is the free electron mass, 
$e$ is the charge of an electron, and $V$ is 
the volume occupied by one formula unit of $\beta$-({\it meso}-DMBEDT-TTF)$_2$PF$_6$.
To evaluate the low-energy spectral weight,
the cut-off energy $\hbar\omega_c$ was adopted 
to be 0.3 eV at which the spectra exhibit  a minimum above $T_c$ [Fig. 2(a)].
Figure 3 shows the temperature variations of $N_{\rm eff}$ obtained using several extrapolation methods for reflectivity spectra. 
Note that difference in the extrapolation methods is negligible in following discussion.
Now two important features are noticed: 
firstly, $N_{\rm eff}$ is increased with lowering temperature above $T_c$.
This indicates that a temperature-dependent transfer energy probably due to a shrinkage of the sample volume 
contributes to the conductive nature in this phase.
This differs from the situation in conventional metals, in which temperature variation of resistivity is mostly 
governed by the reduced scattering rate.
Secondly,
$N_{\rm eff}$ is significantly smaller than unity, 
showing that the low-energy metallic weight is about 10\% 
of what is expected in conventional metals.
We stress that this $N_{\rm eff}$ involves a sizable contribution from high-energy transitions including the dimer peak 
as seen in Fig. 2(a),
and therefore the contribution from conduction electrons should be smaller than the $N_{\rm eff}$ values shown in Fig. 3.
Furthermore,
the magnitudes of $\sigma_1$ and the resulting $N_{\rm eff}$ in $\beta$-(\textit{meso}-DMBEDT-TTF)$_2$PF$_6$ 
are roughly one order smaller than those in $\theta$-(ET)$_2$I$_3$ \cite{Takenaka05} and 
such considerable suppression of low-energy spectral weight has been observed 
in the dimer-Mott insulator $\kappa$-(ET)$_2$Cu[N(CN)$_2$]Cl \cite{Dumm09} and 
the Mott insulating phase of NiS$_{2-x}$Se$_x$ \cite{Perucchi09}.
These results capture an insulating nature at the high-temperature phase
in $\beta$-(\textit{meso}-DMBEDT-TTF)$_2$PF$_6$, and
thus indicate that the charge order transition at $T_c$ is \textit{a transition from dimer-Mott to charge-order phase}.

\begin{figure}[t]
\includegraphics[width=1\linewidth]{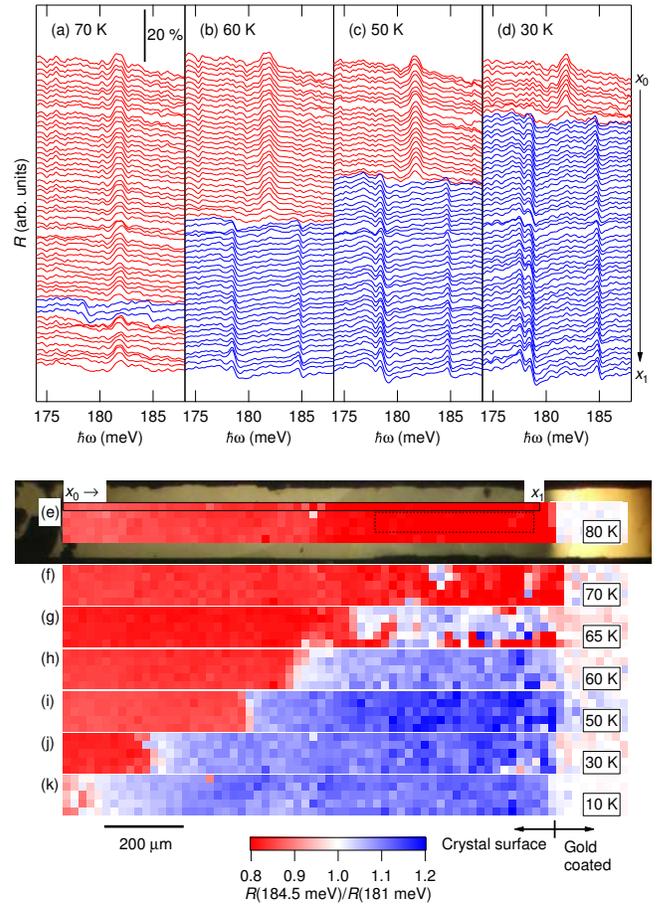}
\caption{(Color)
Spatially-competing charge-order and dimer-Mott insulating states in $\beta$-({\it meso}-DMBEDT-TTF)$_2$PF$_6$.
(a-d) The vibrational molecular modes near 180 meV in the local reflectivity spectra $R(\omega)$ measured with polarization parallel to the {\it c}* axis.
On each panel, 60 local reflectivity spectra, 
which were measured in the region surrounded by solid-line rectangular box 
(from $x_0$ to $x_1$ point) shown in (e), are displayed.
These spectra are separated by 20 $\mu$m and span a total distance of 1.2 mm as represented with the vertical offset.
(e-k) The reflectivity-ratio mapping on the crystal surface of $1400\times 100$ $\mu$m$^2$. 
The color scale shows the reflectivity ratio $R$(184.5 meV)$/R$(181 meV).
The right-side white region is a gold-coated area for measuring the reference spectra.
The dashed-line rectangular box in (e) shows a region in which the large-area reflectivity measurements were performed.}
\end{figure}

Next let us discuss the low-temperature charge-ordered phase.
As seen in Fig. 2(a), the growth of dimer peak
is completely quenched by the formation of charge order below $T_c$,
indicating a competitive nature between dimer-Mott and charge-order phases.
An intriguing question is how these insulating phases compete in real space.
Here we show the positional dependence of local reflectivity spectrum.
Figures 4(a-d) show the vibrational molecular modes near 180 meV 
in the local reflectivity $R(\omega)$ measured at several temperatures.
Each panel displays 60 local reflectivity spectra measured at from $x_0$ to $x_1$ points shown in Fig. 4(e).
As seen in Fig. 4(a), reflectivity spectra are spatially homogeneous at $T_c$: 
almost all spectra possess the single peak near 181 meV (red-color spectra) as expected 
in the dimer-Mott phase [Fig. 2(b)]. 
Charge ordering characterized by the split peak at 178.2 and 184.5 meV emerges 
at a tiny portion inside the crystal as shown by the blue-color spectra.
Meanwhile, the spectra measured below $T_c$ [Figs. 4(b-d)] 
are obviously inhomogeneous and  can be sorted into two groups,
red-color spectra having one single peak originating
from the dimer-Mott insulating state and blue-color spectra 
with split peaks from the charge-ordered state.

Here
we evaluate the reflectivity ratio $R$(184.5 meV)$/R$(181 meV) at each point, 
which gives a relative strength between the charge-ordered state with 184.5-meV peak and 
the dimer-Mott state with 181-meV peak,
and plot its spatial distribution in Figs. 4(e-k).
The red- and blue-color areas indicate the dimer-Mott and the charge-ordered states, respectively.
Note that the large-area reflectivity spectra shown in Figs.~2(a) and (b) were measured in
the rectangular box surrounded by dashed line in Figs. 4(e), 
from which the charge order appears just below $T_c$.
In contrast to spatially-homogeneous spectra above $T_c$, the low-temperature spectra are highly inhomogeneous.
Note that there is negligible temperature gradient inside the crystal \cite{sup}.
The observed inhomogeneity does not originate from the phase separation 
in the hysteresis region near first-order phase transitions 
since the resistive hysteresis loop is closed within 10 K near $T_c$ as shown in the inset of Fig. 3, 
while the observed  inhomogeneity survives far below $T_c$.
Indeed, the present spatial pattern does not depend on the past environment and shows no hysteresis \cite{sup}.
An extrinsic stress effect is also excluded \cite{sup}. 
We note that our results are obtained at ambient pressure,
in sharply contrast to spatial inhomogeneity observed only in pressure \cite{Tanaka08}.

Below $T_c$,  the charge-ordered state gradually invades the dimer-Mott insulating state with lowering temperature,
significantly different from conventional phase transitions occurring at finite temperatures,
at which
a  high-entropy phase at high temperatures is immediately replaced by a low-temperature low-entropy phase. 
In this material,
most surprisingly, the dimer-Mott state survives even at 10 K far below $T_c$.
This experimental fact strongly indicates that
the phase transition in this material is not driven by entropy term in the free energy.
But rather, this transition seems to occur, when the materials parameters reach a critical value through their temperature variation.
We suggest that this type of transition is of quantum nature in the sense that the transition is driven by the materials parameters.
In the present system, it has been found that the interdimer transfer integral is doubly increased from room temperature to 11.5 K \cite{unp}.
Such a considerable change of interdimer integral drives a phase transition from dimer-Mott to antiferro-type electric dipole order \cite{Dayal}.
Now the phase competition expands an inhomogeneous region 
near the border between those two insulating ground states in the parameter space \cite{Burgy01,Dagotto05}.
The present compound may locate near the border at low temperatures.

Here we stress that the observed inhomogeneity well explains previous results. 
While the peak splitting in the local optical spectroscopy is abrupt \cite{Tanaka08}, 
the superlattice intensity of bulk x-ray measurement exhibits a gradual increase with temperature \cite{KimuraJACS}. 
Our results show that the gradual increase originates from the volume-fraction change of charge-ordered state. 
Below $T_c$, the magnetic susceptibility sets to a finite value \cite{unp}, 
indicating the survived dimer-Mott state inside the crystal, well consistent with the present results.

Let us finally discuss  the nucleation mechanism of charge order.
In the previous samples, the superlattice peak intensity and the resistivity are gradually increased below 90 K \cite{KimuraCC,KimuraJACS},
while a recent sample of higher quality shows a sharp increase of resistivity at $T_c$ \cite{unp}.
This difference may originate from the sample quality.
Also, the seeding position of charge order depends on sample \cite{sup}.
Thus we speculate that the seeding is unavoidable crystalline imperfections,
as also indicated from other competing correlated system \cite{Kolb04}. 
Similar nucleation has also been proposed in a ferroelectric relaxor: 
a polar domain is created near the nano-sized chemically ordering regions \cite{Fu09}.
After the nucleation, the charge-ordered state is gradually expanded into the whole region with temperature.
In the present compound, $dT_c(P)/dP$ ($P$: pressure) is negative \cite{Tanaka08}, indicating that 
the volume of charge-ordered state is larger than that of dimer-Mott state.
Thus, the induced charge-ordered state pressurizes the dimer-Mott state near the boundary locally,
leading to $T_c$ reduction in such a boundary region.
This indicates a large energy to move the boundary,
that may cause the observed macroscopic inhomogeneity.

In summary, the infrared measurements on $\beta$-({\it meso}-DMBEDT-TTF)$_2$PF$_6$ reveal
anomalous phase transition phenomena from dimer-Mott to charge-order state.
We suggest a quantum nature of this transition driven by variations of temperature-dependent materials parameters,
which possibly induce a spatial inhomogeneity owing to competitive nature between two insulating states.
This quantum nature might be ubiquitous  among  organic systems which exhibit 
phase transitions at high temperatures, 
where the materials parameters considerably change with temperature.

We thank Y. Nogami, V. Robert, H. Seo, Y. Suzumura, H. Taniguchi and M. Tsuchiizu for fruitful discussion.
The imaging experiments using synchrotron radiation were performed at BL43IR in SPring-8 
with the approvals of JASRI (No. 2011B1221, 2011B1232, 2012A1082, 2012A1141, 2012B1352, 2012B1223).
This work was supported 
by a Grant-in-Aid for Scientific Research on Innovative Areas 
``Molecular Degrees of Freedom'' and 
``Heavy Electrons'' from 
MEXT, Japan.

\newpage
$ $
\newpage

\section{Supplemental Material}

\subsection{Temperature homogeneity}

Figures 5(a) and (b) show the positional dependence of the dimer-peak energy at 80 and 70 K, respectively, 
obtained from the same sample shown in the main manuscript.
Figure 5(c) depicts the positional dependence of the dimer-peak energy in the dotted outlined area shown in Fig. 5(b). 
The dimer-peak energy is spatially homogeneous above 70 K. 
This clearly shows that the sample temperature is also spatially homogeneous above 70 K, 
because the dimer-peak energy exhibits strong temperature dependence at the high-temperature phase 
as seen in Fig. 2(a) in the main manuscript.
In fact, the 80-K and 70-K results in Fig. 5 are explicitly distinguished with each other. 
If there is a temperature gradient inside this sample, a position-dependent dimer peak energy will be obtained as shown by the dashed lines in Fig. 5(c). 
On the other hand, 
our experimental results show no spatial variations of the dimer peak energy. 
This evidences that our sample has negligible temperature gradient.

\subsection{Sample dependence}

In Fig. 6, we show the imaging data measured on other single crystals. 
In the sample \#2, the blue-color charge-order state emerges from the left side,
indicating that the gold deposition for the reference measurement does not affect the observed inhomogeneity.
In the sample \#3, the charge-order state appears from the left and right sides.
These samples show spatial inhomogeneity below 70 K, 
indicating that the inhomogeneity is an intrinsic property in $\beta$-({\it meso}-DMBEDT-TTF)$_2$PF$_6$.

\subsection{Extrinsic stress effect}

In $\kappa$-(ET)$_2$Cu[N(CN)$_2$]Cl, grease coating induces 
a trace of superconductivity due to the extrinsic stress 
originating from the difference of expansion coefficients between samples and grease \cite{S_coat}.
Here we fixed the sample on the cold head by using the carbon paste, 
which may affect the transition temperature to charge ordering 
because the external pressure reduces $T_c$ in
$\beta$-({\it meso}-DMBEDT-TTF)$_2$PF$_6$ \cite{S_Tanaka08}.
To clarify this extrinsic effect, 
we fixed a part of the crystal on the cryostat by using paste, and measured the positional variation of local reflectivity spectrum. 
Figure~7 shows the local reflectivity spectra measured with 50-$\mu$m interval on sample \#4.
The charge order state with split peak was developed from the fixed area, 
clearly showing that 
the survived dimer-Mott state far below $T_c$ does not originate from the stress effect due to the carbon paste.
Moreover, the charge order appears just below $T_c$ in all samples we measured,
excluding such extrinsic stress effect.

\subsection{Hysteresis}

The present spatial pattern does not depend on the past environment and 
shows no hysteresis.
The experiment shown in the main text was done as follows:
the sample was firstly cooled-down to 30 K, and then the mapping data was measured from 30 K to 80 K
with heating.
Then, the sample was cooled-down from 80 K to 50 K, and 50-K data was measured again.
After that, 10-K data was measured.
Figures 8(a) and (b) show the reflectivity-ratio mapping measured at 50 K in the first heating and second cooling processes, respectively.
Similar pattern is found in both results.
Figure 9 displays the reflectivity-ratio mapping measured on the same sample in the main paper 
with fast scanning rate (40 minutes per one imaging).
The results are almost same as those measured with slow scanning rate (3 hours per one imaging) shown in Fig. 4 (main paper), 
indicating that
the sample is essentially in equilibrium at each measurement temperature.

\begin{figure*}[t]
\includegraphics[width=0.75\linewidth]{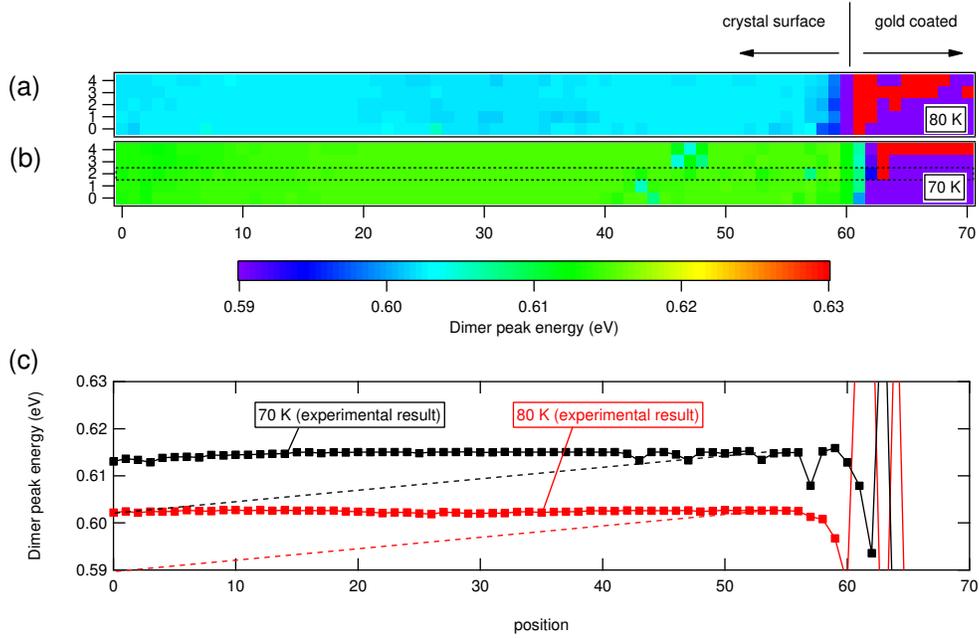}
\caption{
The positional dependence of the dimer-peak energy at (a) 80 K and (b) 70 K  measured on the same sample shown in the main manuscript.
(c) The positional dependence of the dimer-peak energy in the dotted outlined area shown in (b).
The square symbols are the present experimental data, which show no spatial variations.
If there is a temperature gradient in this sample, a position-dependent dimer peak energy will be obtained as shown by the dashed lines. 
The dashed lines are drawn on the assumption that there is 10-K difference between left and right sides.
}
\end{figure*}
\begin{figure*}[t]
\includegraphics[width=0.85\linewidth]{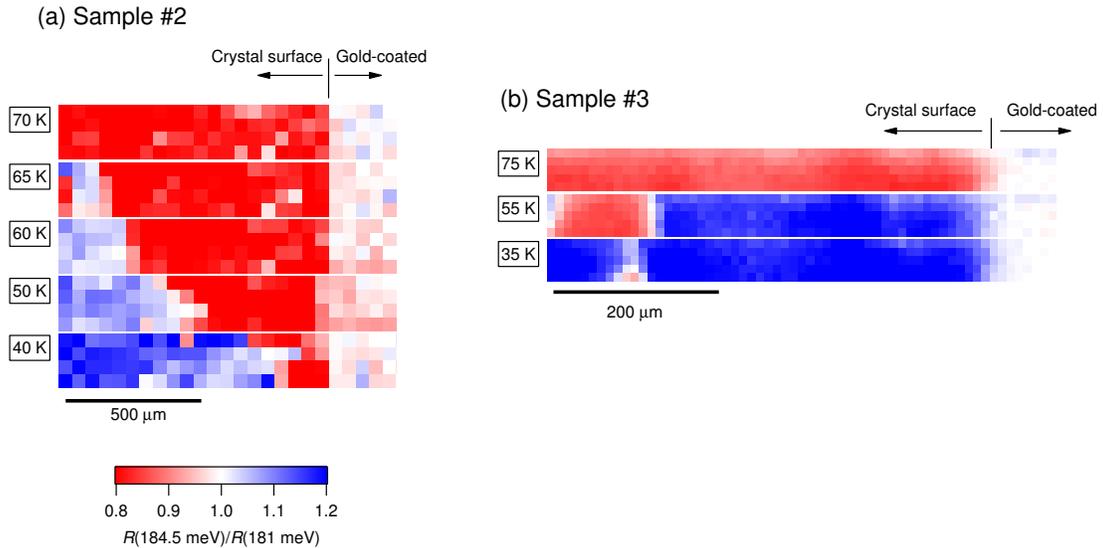}
\caption{
The reflectivity-ratio mapping on (a) sample \#2 and (b) \#3.
}
\end{figure*}
\begin{figure*}[t]
\includegraphics[width=0.45\linewidth]{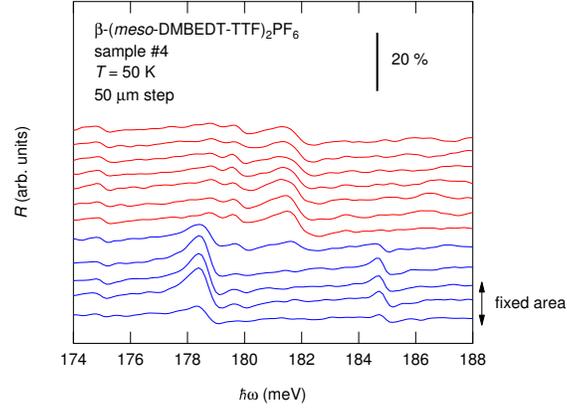}
\caption{
The vibrational molecular modes near 180 meV in the local reflectivity spectra $R(\omega)$ of  sample \#4 
measured with polarization parallel to the {\it c}* axis at $T = 50$ K.
These spectra are separated by 50 $\mu$m and span a total distance of 0.6 mm as represented with the vertical offset.
In this experiment, a part of the crystal is fixed on the cryostat with using a carbon paste.
The fixed area is shown by the arrow.
}
\end{figure*}
\begin{figure*}[t]
\includegraphics[width=0.7\linewidth]{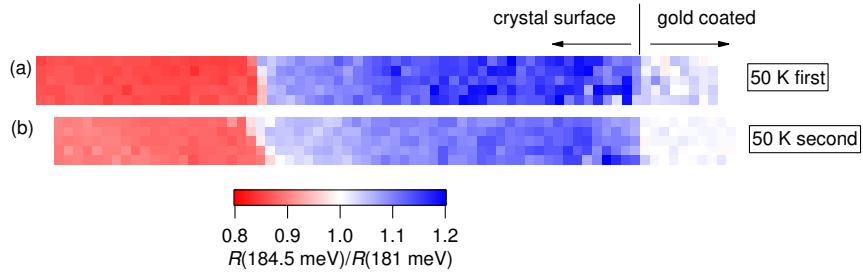}
\caption{
The reflectivity-ratio mapping measured at 50 K  in (a) the first heating and (b) the second cooling process on the same sample in the main paper.
}
\end{figure*}
\begin{figure*}[t]
\includegraphics[width=0.5\linewidth]{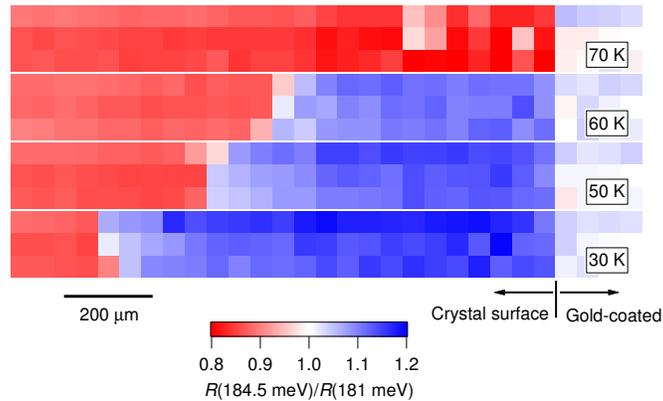}
\caption{
The reflectivity-ratio mapping quickly measured on the same sample in the main paper with coarse spatial resolution (50-$\mu$m interval).
The imaging patterns are almost same as the results measured slowly with fine spatial resolution.}
\end{figure*}

\end{document}